\documentstyle[epsf]{elsart}
\makeatletter
\def\eqalign#1{\null\,\vcenter{\openup\jot\m@th
  \ialign{\strut\hfil$\displaystyle{##}$&$\displaystyle{{}##}$\hfil
      \crcr#1\crcr}}\,}
\def\eqalignleft#1{\null\,\vcenter{\openup\jot\m@th
  \ialign{\strut$\displaystyle{##}$\hfil&$\displaystyle{{}##}$\hfil
      \crcr#1\crcr}}\,}
\makeatother
\begin{document}
\begin{frontmatter}
\title{First LIGO events: binary black holes mergings}

\author{V.~M.~Lipunov, K.~A.~Postnov and M.~E.~Prokhorov}

\address{Moscow State University,
Sternberg Astronomical Institute, 119899 Moscow, Russia\\
e-mail: lipunov@sai.msu.su; pk@sai.msu.su; mike@sai.msu.su\\
fax: +7 (095) 932 88 41}

\begin{abstract}

Based on evolutionary scenarios for binary stellar evolution we study
the merging rates of relativistic binary stars (NS+NS, NS+BH, BH+BH)
under different assumptions of BH formation.  We find the BH+BH merging
rate in the range one per 200,000 -- 500,000 year in a Milky-Way type
galaxy, while  the NS+NS merging rate $R_{ns}$ is approximately 10
times as high, which means that the expected event rate even for high
mean kick velocities of NS up to 400 km/s is at least 30-50 binary NS
mergings per year from within a distance of 200 Mpc.

As typical BH is formed with a mass 3-10 times the NS mass (assumed 1.4
M$_\odot$), the rates obtained imply that the expected detection rate of
binary BH by a LIGO-type gravitational wave detector is 10-100 times
higher than the binary NS merging rate for a wide range of parameters.

\end{abstract}
\end{frontmatter}

\section{Introduction}

The final merging stages of binary relativistic star evolution
containing two compact starts (NS or BH) that merge on a time-scale
shorter than the Hubble time are among the primary targets for
gravitational wave interferometers currently under construction (LIGO,
VIRGO, GEO-600) (Abramovici et al. 1992; Schutz 1996).

It is very important to know the accurate rate of such events, as the
planned LIGO sensitivity will allow detection of NS+NS mergings out to
$\sim 200$ Mpc.  The galactic merging rate of binary NS have constantly
been made over last 20 years and still presents a large controversy
spanning from one per several 1000 yr to several 100000 yr. The
"optimistic" high merging rate have been persistently obtained from
theoretical considerations (Clark et al.  1979, Lipunov et al. 1987,
Narayan et al.  1991, Tutukov \& Yungelson 1993, Lipunov et al. 1995,
Dalton \& Sarazin 1995, Portegies Zwart \& Spreeuw (1996)), whereas the
"ultraconservative" and "realistic" estimates rely upon binary pulsar
statistics only (Phinney 1991, Curran \& Lorimer 1995, van den Heuvel
\& Lorimer 1996) involving as minimum as possible (if any)  theoretical
arguments.

A criticism of the theoretical evolutionary estimates is usually made
with the reference to a large number of poorly determined parameters of
the evolutionary scenario for massive binary systems, such as common
envelope stage efficiency, initial mass ratio distribution,
distribution of the recoil velocity imparted to NS at birth etc.
(Lipunov et al. 1996a).  However, by comparing the results of the
Scenario Machine calculations with other observations (Lipunov et al.
1996a) we may fix some free
parameters (such as the form of the initial mass ratio distribution and
the common envelope efficiency), and then examine the dependence of the
double NS merging rate on one free parameter, say the mean kick
velocity value. In fact, the calculations turn out to be most sensitive
to just the kick velocity (Lipunov et al. 1996b)
as the binary system gets more
chances to be disrupted during supernova explosion, especially when the
recoil velocity becomes higher than the orbital velocity of stars in
the system. This becomes especially important in view of new pulsar velocity
determination (Lyne \& Lorimer 1994) which is indicative of a very
high mean space velocity of $\sim 400-500$ km/s.

The situation is even more poor with binaries containing BH -- no BH+NS
system is known so far (despite a not too pessimistic theoretical
prediction of 1 PSR+BH per 1000 single radiopulsars; Lipunov et al.
1994), hence no "conservative" estimates can be done. Theoretically,
however, putting aside the absolute galactic value of NS+NS and
NS/BH+BH merging rates, we may estimate the ratio of the both,
$N=R_{bh}/R_{ns}$. The important main parameters here are the threshold
mass of a star evolving to a black hole, $M_{cr}$, the mass of the
BH formed during the collapse, $M_{bh}$, and the kick velocity
the BH acquires.

In this paper we calculate this ratio within the
framework of different evolutionary scenarios for massive binary star
evolution for a wide range of parameters. We find that typically
galactic binary BH merging rate $R_{bh}$ is 1-2 orders of magnitude
less than $R_{ns}$, with a smaller difference for high kick velocities
imparted to NS at birth. Nonetheless, since the typical BH mass is 3-10
times higher than that of NS and gravitational waveform's dimensionless
amplitude $h_c$ from a coalescing binary nearly linearly depends upon
mass involved, a GW-detector with a given noise level will be sensitive
to 3-10 times farther BH-systems.  Therefore one may expect a
comparable and even higher number of BH-events than NS-events over the
same observational time, which is of important fundamental character.

Qualitatively, a crude estimate of the ratio of BH-containing binary
merging rate to NS binary merging rate may be done as follows.
Let us assume the mass of BH progenitor to be $35 M_\odot$, which
would correspond roughly to $M_{ms}\sim 60 M_\odot$ on the main sequence
(since according to  the evolutionary scenario, the mass of a star
after mass transfer is $M_{core}\simeq 0.1 M_{ms}^{1.4}$).  On the
other hand, any star with $M_{ms}\ge 10 M_\odot$ evolves to form a NS.
Using the Salpeter mass function ($f(M)\propto M^{-2.35}$), we obtain
that BH formation rate relates to NS formation rate as
$(60/10)^{-1.35}\approx 0.09$.  Extrapolating this logic to binary
BH/NS systems, we might expect $R_{bh}/R_{ns}\sim 1/10$, to a half-order
accuracy. Actually, the situation is complicated by several factors:
the presence of the kick velocity during supernova explosion which may
act more efficiently in the case of NS formation; mass exchange between
the components; distribution by mass ratio, etc. All these factors will
be accounted for in our calculations.

To perform evolutionary calculations, we apply the Monte-Carlo method for
binary stellar evolution studies developed by us over last ten years;
we refer to Lipunov et al. 1996a,b,c for a detailed description of the method and
evolutionary scenarios used.

\section{Parameters of black hole formation}

A black hole  is known to be fully described by three parameters: its
mass $M_{bh}$, angular momentum, and electric charge. For our purposes,
however, only mass is important as it determines the orbital evolution
when the BH enters a close binary system.  Since no exact theory of
stellar-mass BH formation exists, we should somehow parametrize it.
Here we may either fix the initial mass of main-sequence star,
$M_{ms}$, that yields a BH in the end of its evolution, or fix the
threshold pre-supernova mass, $M_*$, that collapses into a BH. The
second parameter is the mass of BH itself, which we assume to be
linearly proportional to the pre-supernova mass: $M_{bh}=k_{bh}M_*$,
$0<k_{bh}\le 1$.  In the case of single stars these two means are fully
equivalent, while when in a close binary with mass exchange between the
components they may be thought to give different results.  Physically,
the latter parametrization ($M_*, k_{bh}$) seems more preferable.  In
fact, we tried both  variants and found them giving only slightly
different figures.

Among other parameters of BH formation in binaries, the kick velocity
imparted to BH during the collapse is the most crucial from the point
of view of binary system evolution. We assume a universal mechanism
giving anisotropic velocity for both neutron stars and black holes,
with BH kick velocity $w_{bh}$ being proportional
to the mass lost during the collapse:
\begin{equation}
w_{bh}=w_{ns}\frac{1-k_{bh}}{1-\frac{M_{OV}}{M_*}}
\end{equation}
where $M_{OV}=2.5$M$_\odot$ is the Oppenheimer-Volkoff limit for NS
mass. This law is chosen  assuming boundary conditions
$w_{bh}=0$ at $k_{bh}=1$ (i.e. when the total mass of the collapsing
star goes into a BH) and $w_{bh}=w_{ns}$ once $M_{bh}=M_{OV}$.

The 3-D kick velocity is assumed to be arbitrarily
oriented in space and to be distributed so as to
fit Lyne-Lorimer pulsars' transverse velocities
(Lyne \& Lorimer, 1994) (see Lipunov et al. 1996ab):

\begin{equation}
f_{LL}(x)\propto \frac{x^{0.19}}{(1+x^{6.72})^{1/2}}
\label{LLkick}
\end{equation}
where $x=w/w_0$, $w_0$ is a parameter, it fits well the  Lyne \& Lorimer's
2D-distribution at $w_0=400$~ km/s.

\section{The detection rate of BH mergings versus NS mergings}

\begin{figure}
\centerline{%
\epsfysize=0.5\hsize%
\epsfbox{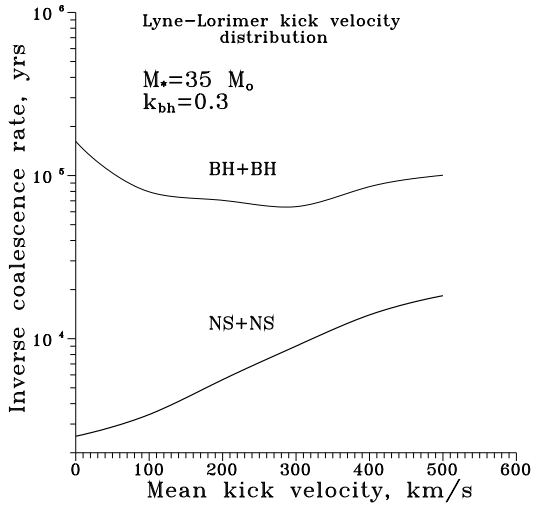}}
\caption{The galactic merging rate of NS+NS  and BH+BH binaries
for Lyne-Lorimer kick velocity distribution as a function
of the mean kick velocity assuming BH formation parameters
$M_*=35 M_\odot, k_{bh}=0.3$}
\label{rate}
\end{figure}

Fig. \ref{rate} shows the relativistic compact binaries' merging rates
as a function of the mean kick velocity assuming Lyne-Lorimer kick
velocity distribution (\ref{LLkick}). In this variant, BH parameters
were chosen $M_*=35 M_\odot, k_{bh}=0.3$, i.e. BH with masses
$M_{bh}>11.5 M_\odot$ were formed during evolution.  From Fig.
\ref{rate} we see that the theoretical expectation for the NS+NS
merging rate in a model spiral galaxy with a typical mass of $10^{11}$
M$_\odot$ lie within the range from $\sim 3\times 10^{-4}$ yr$^{-1}$ to
$\sim 10^{-5}$ yr$^{-1}$, depending on the assumed mean kick velocity
and the shape of its distribution. For Lyne \& Lorimer kick velocity
law with the mean value of 400 km/s, we obtain $R_{NS+NS}\approx
5\times 10^{-5}$ yr$^{-1}$.  To obtain the merging rate from a given
volume $V$ in the Universe, the galactic rate should be scaled, for
example, using Phinney's formula (Phinney 1991): ${\cal R} \approx 0.01
h_{100} R\times V$.

Two details from Fig. \ref{rate} are worth noting: 1) the binding
effect at small kick velocities and 2) the smaller effect of high kicks
on the BH+BH rate.  The first fact is qualitatively clear: a high kick
leads to the system disruption; however, if the system is survived the
explosion, its orbit would have a periastron always smaller than in the
case without kick. During the subsequent tidal circularization a closer
binary system will form which will spend less time prior to the
merging.  The binding effect of small recoil velocities is very
pronounced in the case of binary BH. At higher kicks their merging rate
decreases slower due to higher masses of the components.

We also present the distribution of merging neutron stars by their ages
(the time from the birth to merging) for ``standard'' scenario
parameters (see the Appendix) and Lyne-Lorimer kick velocity
distribution with $w=400$ km/s. Clearly, the characteristic ages of
merging neutron stars range from $10^{8}$ to $10^{10}$ years and the
vast majority of all merging neutron stars are much older than
charatceristic pulsar age of $5\times 10^6$ years.  This is important
for understanding the difference between evolutionary and
pulsar-statistics-based estimates of NS+NS merging rates (see the
Discussion).

\begin{figure}
\centerline{%
\epsfysize=0.5\hsize%
\epsfbox{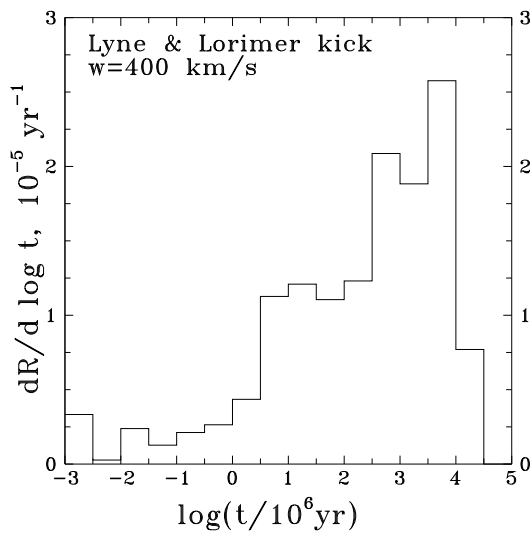}}
\caption{The galactic birthrate of merging NS
with different ages (the time from NS formation to merging)
for Lyne-Lorimer kick velocity distribution and $w=400$ km/s.}
\label{ages}
\end{figure}

Since the characteristic dimensionless strain metric
amplitude from a merging binary system, $h_c$, scales as $M^{5/6}/r$,
where $M$ is a characteristic mass of binary companions and $r$ is a
distance to the source (Abramovici al. 1992)), the number
of events registered by the detector scales as
\[
\label{rates}
\frac{N_{bh}}{N_{ns}}\approx\left(\frac{R_{bh}}{R_{ns}}\right)
                \left(\frac{M_{bh}}{M_{ns}}\right)^{15/6}
\]

Clearly, the second factor in this expression may overcome the
first one (which is of order 0.1-0.01) if typical BH masses
is several times as high as the NS mass.

\begin{figure}
\centerline{%
\epsfysize=0.5\hsize%
\epsfbox{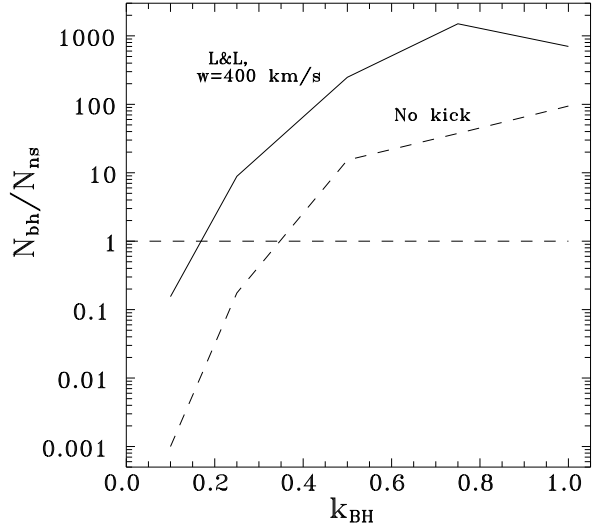}}
\caption{The ratio of BH+BH to NS+NS merging events
as expected to be
detected by a gravitational wave detector with a given sensitivity, as
a function of
parameter $k_{bh}$ (with $M_*=35 M_\odot$)
assuming Lyne-Lorimer kick velocity distribution
with $w_0=400$ km/s (solid curve)
and without kick (dashed curve)}
\label{ratio}
\end{figure}

Fig. \ref{ratio} demonstrates the relative detection rate of
coalescing BH and NS  binaries obtained by a gravitational
wave detector with given sensitivity.
Lyne-Lorimer kick velocity distribution with $w_0=400$ km/s were assumed.
Here we fixed $M_*=35 M_\odot$ and varied $k_{bh}$.
Other scenario parameters (such as
the form of the initial mass ratio distribution and
common envelope efficiency) only
slightly affect the results. It is seen that the ratio of the expected
detection rates for
merging BH/NS binaries (\ref{rates}) may well exceed unity for a wide
range of parameters.

\section{Discussion and conclusions}

The NS+NS merging rates obtained by us is an order of magnitude higher
than those obtained by Portegies Zwart and Spreeuw (1996). Whereas
we both use the same mean kick velocity $w=400$ km/s, the shapes
of kick velocity distribution are totally different: Portegies Zwart
and Spreeuw (1996) use a Maxwellian distribution which produces a
strong deficit of slow pulsars relative to the observed distribution
(Lyne \& Lorimer 1994). In our other paper (see Lipunov et al. 1996b)
we also perform calculations for the Maxwellian
kick velocity distributions and find full agreement with
calculations of Portegies Zwart and Spreeuw. We also showed
(Lipunov et al 1996a, Fig. 9-10) that
the Maxwellian distribution with $w=400$ km/s does not describe
the observed pulsar velocity distribution at low velocities and,
moreover, strongly contradict to the binary pulsar statistics in
general.

Our galactic NS+NS merging rate ($5\times 10^{-5}$ yr$^{-1}$)
is also exceeds the so-called
"Bailes upper limit" to the NS+NS birthrate in the Galaxy
$\sim 10^{-5}$ yr$^{-1}$ (Bailes, 1996).
Indeed, following Bailes, the fraction of "normal"
pulsars in NS+NS binaries which potentially could merge during
the Hubble time among the total number of
known pulsars is less than $\sim 1/700$. Multiplying this by
the birthrate of "normal" pulsars $1/125$ yr${-1}$ we get
this limit $10^{-5}$ yr$^{-1}$. Our value is 5 times higher
for Lyne-Lorimer kick velocity distribution at $w=400$ km/s.
First we note that the accuracy of Bailes limit
is a half-order at best; in addition, the pulsar birthrate
1/125 yr$^{-1}$ is at least 5 times lower than the birthrate
of massive stars ($>10 M_\odot$) in our Galaxy according to
Salpeter mass function (once per 25 years) which
produce neutron stars in the framework of the modern evolutionary
scenario. This discrepancy
may be decreased if we take into account the uncertainty in
the pulsar beaming factor and, which may be more important,
the still-present uncertainty in pulsar distance scale
which influences the estimate of the total galactic number of
pulsars and hence their birthrate.

On the other hand, the ``best guess'' observational limit for NS+NS
merging rate (Phinney 1991; van den Heuvel and Lorimer 1996) yields
$\sim 8\times 10^{-6}$ yr$^{-1}$ which is an order of magnitude smaller
than our calculated rate.  Possible explanations to this discrepancy
could be that many binary NS are born in close orbits and merge in a
short time, as was suggested by van den Heuvel (1992) (see also Tutukov
\& Yungelson (1993); note, however, that our calculations and Fig. 1 of
Tutukov and Yungelson's paper show that  binary NS+NS with time before
coalescence less than $10^7$ years contributes only $\sim 10-20\%$ to
the total galactic merging rate).  We specially calculated the
distribution of merging neutron stars by their ages (see Fig.
\ref{ages}).  It is seen that the mean age of merging NS is about
$10^8$ years; the fraction of merging NS with ages shorter than
$5\times 10^6$ years is $15\%$; so the observational underestimation is
mainly due to a large portion of relativistic binaries containing no
pulsar (ejecting neutron star) and thus being unobservable by
traditional radioastronomical means (Lipunov et al. 1996b).

The example calculations presented above clearly show that the expected
detection rate of mergings with BH may turn out to be much higher than
those with NS. In view of the great fundamental importance of this
finding, the question arises how stable is this result against changing
of other (numerous) parameters? The parameters used in evolutionary
calculations may be subdivided into several groups:

(1) parameters of BH formation (such as progenitor's mass, BH mass,
kick velocity during the collapse);

(2) parameters of binary evolutionary scenario (such as initial mass
ratio distribution, mass transfer treatment, common envelope
efficiency, kick velocity distribution shape, etc.);

(3) general parameters of stellar evolution (mass loss rate at
different stage, convection treatment, etc.).

In the present paper we focused on BH formation parameters fixing others
on the grounds discussing elsewhere (see Lipunov et al. 1996a). In a
separate paper (Lipunov et al. 1996b) we study the influence of the most
important of them (the common envelope efficiency, the initial mass
ratio distribution, the shape of the kick velocity distribution, mass
loss by single stars), and come to essentially the same conclusion: the
expected number of binary BH mergings exceeds that of NS merging for a
wide range of parameters!

This urges studies on possibly exact calculation of GW-waveforms
emitted during BH coalescences to obtain maximum information on this
fundamental and probably the most spectacular natural phenomenon.

\begin{ack}

The authors acknowledge Prof. Ed van den Heuvel for encouraging
discussions.  The work was partially supported by the INTAS grant
93-3364, grant of Russian Fund for Basic Research No 95-02-06053a.
\end{ack}

\section{Appendix: The evolutionary scenario model}

Monte-Carlo simulations of binary evolution allows one to study the
evolution of a large ensemble of binaries  and to estimate the number
of binaries at different evolutionary stages.  This method has become
popular over last ten years ( Kornilov \& Lipunov 1984; Dewey \& Cordes
1987; Bailes 1989; for another applications of Monte-Carlo simulations
see de Kool 1992; Tutukov \& Yungelson 1993; Pols \& Marinus 1994).

For modeling binary evolution, we use the ``Scenario Machine'', a
computer code that includes a modern scenario of binary evolution (for
a review, see van den Heuvel (1994)) and takes into account the
influence of magnetic field of compact objects on their observational
appearance. A detailed description of the computational techniques and
input assumptions is summarized elsewhere (Lipunov et al. 1996a),
and here we list
only basic parameters and initial distributions.

\subsection{Initial binary parameters}

The initial  parameters determining binary evolution are: the
mass of the primary ZAMS component, $M_1$; the binary mass ratio,
$q=M_2/M_1<1$; the orbital separation, $a$. We assume zero initial
eccentricity.

The distribution of initial binaries
over orbital separations is known from observations
(Abt 1983):
\begin{equation}
\eqalign{&f(\log a) ={\rm const}\,,\cr
&\max~\{10~ {\rm R}_\odot,~\hbox{Roche Lobe}~
(M_1)\} < \log a < 10^4~{\rm R}_\odot.
\cr}
\end{equation}

The initial mass ratio distribution in binaries, being very crucial for
overall evolution of a particular binary system (Trimble 1983),
has not yet been derived
somehow reliably from observations due to a number of selection
effects.  A `zero assumption' usually made is that the mass ratio
distribution has a flat shape, i.e. the high mass ratio binaries
are formed as frequently as those with equal masses (e.g. van den Heuvel 1994).
Ignoring the real distribution, we parametrized it by a power
law, assuming the primary mass distribution to obey the Salpeter mass
function (Salpeter 1955):
\begin{equation}
\eqalignleft{
f(M_1) \propto  M_1^{-2.35}\,,\quad &10~{\rm M}_\odot < M_1< 120~{\rm M}_\odot\,;
\cr f(q)   \propto  q^{\alpha_q}\,,\quad&q\equiv M_2/M_1<1 \,;   \cr }
\end{equation}

A comparison of the observed X-ray source statistics with the
predictions of the current evolutionary scenarios indicates
(Lipunov et al. 1996a)
that the initial mass ratio should be strongly centered around unity,
($\alpha_q\sim 2$). Of course, this is not a unique way of
approximating the initial binary mass ratio (see e.g. Tout (1991)).
However, from the point of view of binary NS merging rate,
this parameter affects the results much less
than the kick velocity. In the present paper, we use both
$\alpha_q=2$ and $\alpha_q=0$.

\subsection{Initial parameters of compact stars}

We are interested in binary NS or NS+BH systems,
so it is enough to trace
evolution of binaries with primary masses $M_1>10 M_{\odot}$ which are
capable of producing NS and BH in the end of evolution.
The secondary component can have a mass
from the whole range of stellar masses $0.1 M_\odot<M_2<120 M_\odot$.

We consider a NS with a mass of $1.4~ {\rm M}_{\odot}$ to be
a result of the collapse of a star with the core mass prior to the
collapse $M_*\sim (2.5-35)~{\rm M}_{\odot}$.  This corresponds to an
initial mass range $\sim (10 - 60)~{\rm M}_{\odot}$, considering
that a massive star can loose more than $\sim (10-20)\%$ of its
initial mass during the evolution with a strong stellar wind (de Jager
1980).

The magnetic field of a rotating NS largely defines the evolutionary
stage the star would have in a binary system (Schwartzman 1970;
Davidson \& Ostriker 1973; Illarionov \& Sunyaev 1975).  We use a
general classification scheme for magnetized objects elaborated by
Lipunov (1992).

Briefly, the evolutionary stage of a rotating magnetized NS  in a
binary system depends on the star's spin period $P$ (or spin frequency
$\omega=2\pi/P$), its magnetic field strength $B$ (or, equivalently,
magnetic dipole moment $\mu=BR^3/2$, where $R$ is the NS radius), and
the physical parameters of the surrounding  plasma (such as density
$\rho$ and sound velocity $v_s$) supplied by the secondary star.  The
latter, in turn, could be a normal optical main sequence (MS) star, or
red giant, or another compact star). In terms of the Lipunov's
formalism, the NS evolutionary stage is determined by one or another
inequality between the following characteristic radii: the light
cylinder radius of the NS, $R_l=c/\omega$ ($c$ is the speed of light);
the corotation radius, $R_c=(GM/\omega^2)^{1/3}$; the gravitational
capture radius, $R_G=2GM/v^2$ (where $G$ is the Newtonian gravitational
constant and $v$ is the NS velocity relative to the surrounding
plasma); and the stopping radius $R_{stop}$. The latter is a
characteristic distance at which the ram pressure of the accreting
matter matches either the NS magnetosphere pressure (this radius is
called Alfven radius, $R_A$) or the pressure of relativistic particles
ejected by the rotating magnetized NS (this radius is called
Schwartzman radius, $R_{Sh}$).  For instance, if $R_l>R_G$  then the NS
is in the ejector stage (E-stage) and can be observed as a radiopulsar;
if $R_c<R_A<R_G$, then so-called propeller regime is established
(Illarionov \& Sunyaev 1975) and the matter is expelled by the rotating
magnetosphere; if $R_A<R_c<R_G$, we deal with an accreting NS
(A-stage), etc. These inequalities can easily be translated into
relationships between the spin period $P$ and some critical period that
depends on $\mu$, the orbital parameters, and accretion rate $\dot M$
(the latter relates $v$, $v_s$, $\rho$, and the binary's major semiaxis
$a$ via the continuity equation). Thus, the evolution of a NS in a
binary system is essentially reduced to the NS spin evolution
$\omega(t)$, which, in turn, is determined by the evolution of the
secondary component and orbital separation $a(t)$. Typically, a single
NS embedded into the interstellar medium evolves as $E\to P\to A$
(for details, see Lipunov \& Popov 1995). For a NS in a binary, the
evolution complicates as the secondary star evolves: for example, $E\to
P\to A\to E$ (recycling), etc.

When the secondary component in a binary overfills its Roche lobe, the
rate of accretion onto the compact star can reach the value
corresponding to the
Eddington luminosity $L_{Edd}\simeq 10^{38}~(M/{\rm M}_\odot)$
\hbox{erg/s}\, at the $R_{stop}$; then a supercritical regime sets in
(not only superaccretors but superpropellers and superejectors can
exist as well; see Lipunov 1992).

If a BH is formed in due course of the evolution, it can only appear as
an accreting or superaccreting X-ray source; other very interesting
stages such as BH + radiopulsar which may constitute a notable fraction
of all binary pulsars after a starburst are considered in Lipunov et
al. (1994).

The initial distribution of magnetic fields of NS is another
important parameter of the model.  This cannot be taken from studying
pulsar magnetic field (clearly, pulsars with highest and lowest
fields are difficult to observe).  In the present calculations we
assume a flat distribution for dipole magnetic moments of newborn NS %
\begin{eqnarray} f(\log\mu)=~{\rm const}\,,~~ 10^{28} \le \mu \le
      10^{32}~ \hbox{G cm}^3\,,
\end{eqnarray}
and the initial rotational period of the NS is assumed to be $1$~ms.

The computations were made under different assumptions about the
NS magnetic field decay, taken in an exponential form,
$\mu(t)\propto \exp(-t/tau)$, where $\tau$ is the characteristic decay
time  of $10^7-10^8$ year.  The field is assumed to
stop decaying below a minimum value of $10^9$ G (van den Heuvel et al.
1986). No accretion-induced magnetic field decay
is assumed.

A radiopulsar was assumed to be turned ``on'' until its period $P$ has
reached a ``death-line'' value defined from the relation
$\mu_{30}/P_{death}^2=0.4$, where $\mu_{30}$ is the dipole magnetic
moment in units of $10^{30}$ G~cm$^3$, and $P$ is taken in seconds.

The mass limit for NS (the Oppenheimer-Volkoff limit) is $M_{OV}=2.5~
{\rm M}_\odot$, which corresponds to a hard equation of state of the NS
matter.  The most massive stars are assumed to collapse into a BH once
their mass before the collapse is $M>M_{cr}=35~ {\rm M}_\odot$ (which
would correspond to an initial mass of the ZAMS star $\sim 60~ {\rm
M}_\odot$ since a substantial mass loss due to a strong stellar wind
occurs for the most massive stars).  The BH mass is calculated as
$M_{bh}=k_{bh}M_{cr}$, where the parameter $k_{bh}$ is taken to be 0.3,
as follows from the studies of binary NS+BH (Lipunov et al.  1994).

\subsection{Other parameters of the evolutionary scenario}

The fate of a binary star during evolution mainly depends on the
initial masses of the components and their orbital separation.  The
mass loss and kick velocity are the processes leading to the binary
system disruption; however, there are a number of processes connected
with the orbital momentum losses tending to bound the binary (e.g.,
gravitational radiation, magnetic stellar wind).

\subsubsection{Common envelope stage}

We consider stars with a constant (solar) chemical composition.  The
process of mass transfer between the binary components is treated
according to the prescription given in van den Heuvel (1994) (see Lipunov et al.
(1996a) for more detail). The non-conservativeness of the mass transfer
is treated via ``isotropic re-emission'' mode (Bhattacharya \& van den
Heuvel 1991).  If the rate of accretion from one star to another is
sufficiently high (e.g. the mass transfer occurs on a timescale 10
times shorter than the thermal Kelvin-Helmholz time for the normal
companion), or the compact object is engulfed by a giant companion, the
common envelope (CE) stage of the binary evolution  can set in (see
Paczy\'nski 1976; van den Heuvel 1983).

During the CE stage, an effective spiral-in of the binary components
occurs.  This complicated process is not fully understood as yet, so we
use the conventional energy consideration to find the binary system
characteristics after the CE stage by introducing a parameter
$\alpha_{CE}$ that measures what fraction of the system's orbital
energy goes, between the beginning and the end of the spiralling-in
process, into the binding energy (gravitational minus thermal) of the
ejected common envelope. Thus,
\begin{equation}
\alpha_{CE}\left({GM_aM_c\over 2a_f}-{GM_aM_d\over 2a_i}\right) = {GM_d
    \left(M_d-M_c\right) \over R_d}\,,
\end{equation}
where $M_c$ is the mass of the core of the mass loosing star of initial
mass $M_d$ and radius $R_d$ (which is simply a function of the initial
separation $a_i$ and the initial mass ratio $M_a/M_d$), and no
substantial mass growth for the accretor is assumed (see, however,
Chevalier 1993).  The less $\alpha_{CE}$, the closer becomes
binary after the CE stage. This parameter is poorly known
and we varied it from 0.5 to 10  during calculations.

\subsubsection{High and low mass-loss scenario form massive star
evolution}

A very important parameter of the evolutionary scenario is
the stellar wind mass loss effective for massive stars.
No consensus on how stellar wind mass loss occurs in
massive stars exist.
So in the spirit of our scenario approach
we use two "extreme", in a sense, cases. The "low mass-loss"
scenario  treats
the stellar wind from a massive star of luminosity $L$
according to de Jager's prescription
\[
\dot M \propto \frac{L}{cv_\infty}
\]
where $c$ and $v_\infty$ are the speed of light and of the stellar
wind at infinity, respectively. This leads to at most 30 per cent
mass loss for most massive stars.

The "high stellar wind mass-loss" scenario
uses calculations of single star evolution by Schaller et al. (1992).
According to these calculations, a massive star lose most of its mass by
stellar wind down to 8-10 $M_\odot$ before the collapse,
practically independently on its initial mass.
In this case we assume the same mechanism for
BH formation as for the "low mass-loss" scenario, but
only one parameter $k_{bh}$ remains ($M_{cr}$ is taken from
evolutionary tracks).
Masses
of BH formed within the framework of the high mass-loss
scenario are thus always less than or about of 8 M$_\odot$.

So far we are unable to choose between the two scenarios; however,
recently reported observations of a very massive WR star of 72 M$_\odot$
(Rauw et al. 1996) cast some doubts on very high mass-loss scenario or
may imply that different mechanisms drive stellar wind mass loss.

{}

\end{document}